\documentclass[prl, reprint]{revtex4-1}
\usepackage{amsmath}
\usepackage{graphicx}
\usepackage{graphicx}
\usepackage{amsmath}
\usepackage{amssymb}
\usepackage{url}
\usepackage{bm}		
\usepackage{color}	
\usepackage{textcomp}				

\begin{document}


\title{Localization and time-reversal of light through dynamic modulation} 



\author{Momchil Minkov}

\author{Shanhui Fan}
\email[]{shanhui@stanford.edu}
\affiliation{Department of Electrical Engineering, and Ginzton Laboratory, Stanford University, Stanford, CA 94305, USA}

\date{\today}

\begin{abstract}
We study the dynamics of a waveguide made of coupled resonators with a sinusoidal modulation of the resonance frequencies. We present a modulation scheme that achieves complete dynamic localization and is experimentally suitable for optical cavities. Furthermore, we highlight the importance of the way the modulation is turned on and off. One striking consequence is that, while a state returns to its starting amplitude at the end of every cycle, it can also be fully time-reversed if the modulation is turned off mid-cycle. Finally, we show that localization is always achieved when the modulation envelope is adiabatic with respect to the oscillation frequency. The results are experimentally feasible using existing integrated photonic technologies, and are relevant to applications like tunable delay lines, dispersion compensation, and imaging.
\end{abstract}

\maketitle


The localization of classical and quantum waves has been extensively researched since it was first studied by Anderson in disordered systems \cite{Anderson1958, Kramer1993}. Localization can also occur in regular but time-dependent potentials -- for example when a suitably-designed, time-periodic electric field is driving a charged particle on a lattice \cite{Dunlap1986}. This \textit{dynamic localization} effect, in which the wave-function returns to its starting value at the end of every cycle, has been the object of a number of theoretical studies \cite{Dignam2002, Creffield2010, Tsuji2011}, and has also been experimentally demonstrated using cold atom gases \cite{Fischer1999, Eckardt2009}. An analogous effect has also been studied for photons in a lattice of coupled waveguides \cite{Longhi2005, Longhi2006, Szameit2009, Szameit2010, Joushaghani2012}, in which the role of time is taken by the spatial coordinate in the propagation direction, and the localization analogue is the suppression of diffraction of a propagating beam.

True dynamic localization of light (i.e. in time as opposed to space) is the ultimate realization of slow light, as it corresponds to zero group velocity -- and is thus interesting for applications such as optical buffering \cite{Krauss2008, Baba2008}. Here, we discuss how dynamic localization can be implemented in a coupled-cavity waveguide (CCW) \cite{Yariv1999} using a sinusoidal temporal modulation of the resonance frequencies of the constituent cavities. Unlike the only previous demonstration of localization in a CCW \cite{Yuan2015}, which requires a modulation of the intra-cavity coupling constants, our approach is straightforward to implement using standard electro-optic modulators \citep{Xu2005, Reed2010}. In addition, we also discuss the significance of the \textit{micromotion} associated to the dynamic localization, i.e. the evolution of the state at times that are not integer multiples of the modulation period. This has been largely ignored in previous works, but can in fact produce significant effects. A striking illustration is the fact that a wave-packet can be fully time-reversed if the localizing modulation is stopped at a precise moment. Such a time-reversal operation is important for dispersion compensation \cite{Yariv1979} and imaging through complex media \cite{Mosk2012}.

We start our discussion with the CCW illustrated in Fig. \ref{fig1}(a). Following Refs. \cite{Haus1984, Minkov2017}, we write the coupled-mode theory equation of the system in terms of the electric field amplitude $\alpha_n$ in each resonator,
\begin{equation}
\frac{\partial \alpha_n}{\partial t} = i(\omega_0 + \omega_n(t)) \alpha_n + i\sum_m J_{nm} \alpha_m,
\label{eqn:cm}
\end{equation}
where $J_{nm} = J$ for $m = n \pm 1$, and $0$ otherwise. We focus specifically on a modulation of the form $\omega_n(t) = A_n \cos(\Omega t + \varphi_n)$. Under the gauge transformation  $\beta_n(t) = \alpha_n(t) \times \exp\left(-i\omega_0 t -i \int_0^t \omega_n(t') \mathrm{d}t'\right)$, the time-dependence of $\omega_n(t)$ translates into a gauge field, appearing as a complex time-dependence of the coupling coefficients. We derive this explicitly in the Supplementary Information \cite{SM}, and we study perturbatively the dynamics of $\beta_n(t)$ in the high-frequency limit of $J \ll \Omega$. In this limit of small modulation period $T = 2\pi/\Omega$, the effective time-independent equations of motion governing the \textit{stroboscopic} time evolution at times $qT$, $q \in \mathcal{Z}$ have the form:

\begin{figure}
\centering
\includegraphics[width = 0.45\textwidth, trim = 0in 0in 0in 0in, clip = true]{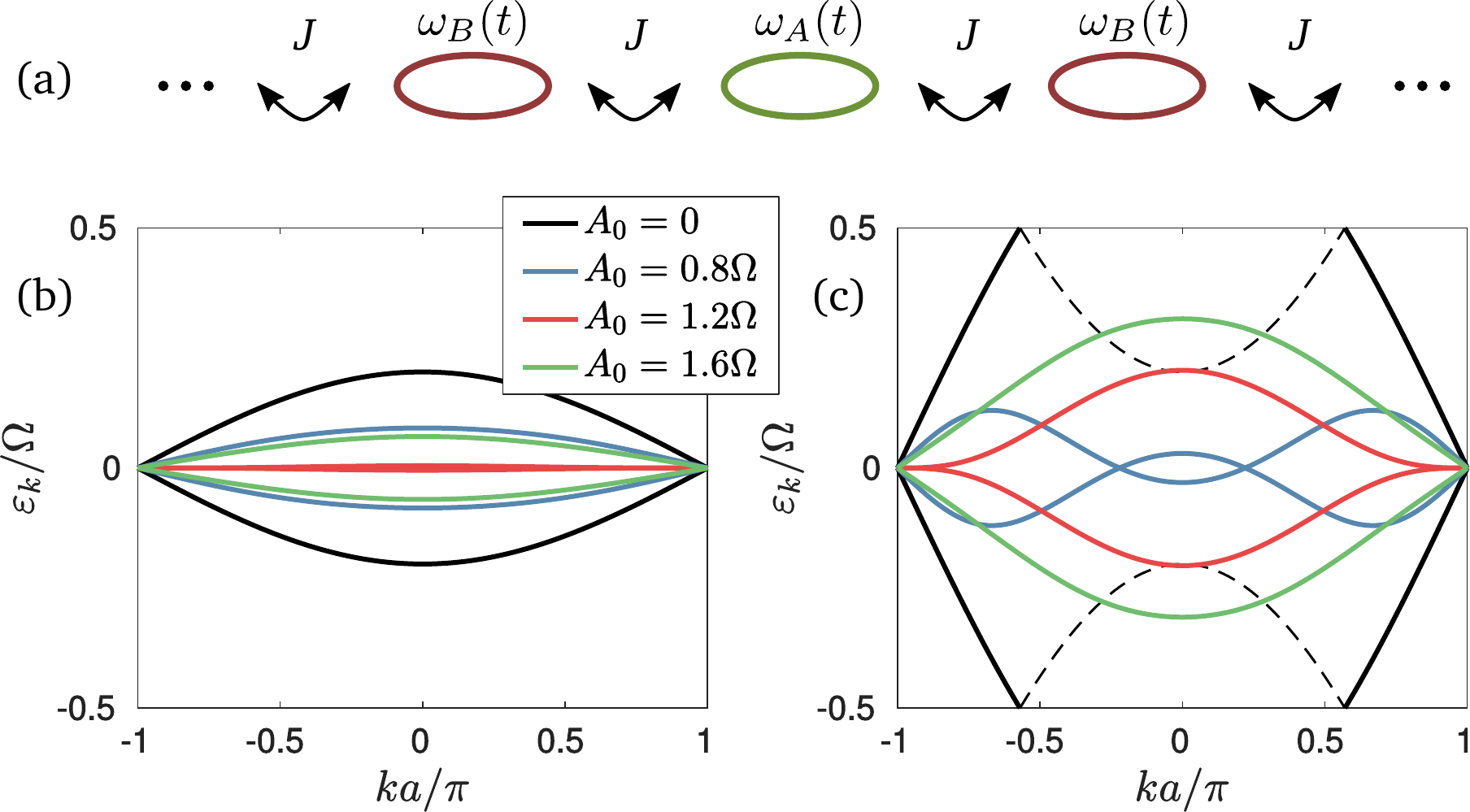}%
 \caption{(a): Schematic of the system of coupled resonators with a coupling constant $J$ and a time-dependent resonance frequency $\omega_0 + \omega_n(t)$. We set the distance between resonators to $1$, and furthermore consider a spatially periodic modulation with period $a = 2$. (b): Band diagram of the Floquet quasi-energies for $\omega_n(t) = A_0\cos(\Omega t + n\pi)$, for $J = 0.1\Omega$ and for several different values of $A_0$. (c): Same as (b), for $J = 0.4\Omega$. The dashed lines illustrate the $A_0 = 0$ dispersion folded into the temporal Brillouin zone.}
\label{fig1}
\end{figure}
\begin{equation}
\frac{\partial \beta_n}{\partial t} = i \sum_{m} \left(J^{(1)}_{nm} + J^{(2)}_{nm} \right) \beta_m,
\end{equation}
The first- and second-order effective coupling constants are given by 
\begin{align}
&J^{(1)}_{nm} = J_{nm} \mathcal{J}_0 \left(\rho_{nm}\right) ,
 \label{eqn:effective_J} \\ 
&J^{(2)}_{nm} = -2i \sum_p J_{np} J_{pm} \times \\ \nonumber & \quad \sum_{k > 0} \frac{(-1)^k}{k\Omega} \mathcal{J}_k(\rho_{np})\mathcal{J}_k(\rho_{pm}) \sin(k(\varphi_{np} - \varphi_{pm})), \end{align}
where 
\begin{align} \rho_{np} e^{i\varphi_{np}} \equiv \left(\frac{A_n}{\Omega}  e^{i \varphi_n} - \frac{A_p}{\Omega} e^{i \varphi_p} \right), 
\label{rhophi}
\end{align}
for arbitrary integer $n$, $p$, and $\mathcal{J}_n(x)$ is the $n$-th order Bessel functions of the first kind. This result can also be derived from the high-frequency approximation of the time-dependent Hamiltonian formalism used in cold-atom systems \cite{Rahav2003, Goldman2014, Jotzu2014, Eckardt2015}. Note that the second-order term $J^{(2)}_{nm}$ is purely imaginary, and hence non-reciprocal, and can be used for applications such as optical isolation \cite{Fang2012}, or as a basis for a photonic topological insulator \cite{Minkov2016}. This is however beyond the scope of the current work -- our interest is instead light localization, for which the requirement is simply that the effective couplings go to zero. We note that the original system of Ref. \cite{Dunlap1986} is straightforward to replicate with a modulation of the form $A_n \cos(\Omega t) = nA_0 \cos(\Omega t)$. However, this choice requires a spatial gradient in the amplitude, which is difficult for optical resonators. In our scheme, by setting $\omega_n = A_0 \cos(\Omega t + n\pi)$, the resulting effective couplings are $J^{(1)} = J \mathcal{J}_0(2A_0/\Omega)$, $J^{(2)} = 0$, and so localization occurs when $2A_0/\Omega$ is a root of $\mathcal{J}_0(x)$. This constant-amplitude scheme is much more feasible for optical cavities, but, interestingly, it is difficult for electronic systems, as it corresponds to an electric field with microscopic spatial periodicity. 

The modulation $A_0 \cos(\Omega t + n\pi)$ is periodic both in space and in time, with a spatial period $a = 2$ (Fig. \ref{fig1}(a)). The Floquet-Bloch theorem thus guarantees that any solution is a superposition of states of the form
\begin{equation}
\alpha_{k, j}(x, t) = e^{ikx} e^{-i\omega_0 t - i\varepsilon_j(k)t} u_{k, j}(x, t),
\end{equation}
where $k \in [-\pi/a, \pi/a]$ is the Bloch momentum, $\varepsilon_j(k) \in [-\Omega/2, \Omega/2]$ is the quasi-energy, $j$ labels different Floquet bands, and $u_{k, j}(x, t)$ is periodic both in space and time, with spatial period $a$ and temporal period $T$. For our constant-amplitude modulation scheme, we compute the quasi-energy band structure $\epsilon_j(k)$, for various parameters $J$ and $A_0$. In Fig. \ref{fig1}(b), this is shown for $J = 0.1\Omega$. With no modulation, the dispersion is simply given by $\varepsilon_j(k) = \pm 2J\cos(ka/2)$, as for a regular CCW with two cavities included in the elementary cell. Furthermore, since $J/\Omega \ll 1$, in the presence of modulation the bands are well-described by the perturbative prediction, namely $\varepsilon_j(k) = \pm 2J^{(1)} \cos(k a/2)$. For $A_0 = 1.2\Omega$ in particular, $J^{(1)} \approx 0$, and the corresponding band appears flat, as expected for dynamic localization. This observation is related to the light-stopping scheme of Ref. \cite{Yanik2004a}, where the actual photonic dispersion (as opposed to the Floquet quasi-energy band) is adiabatically flattened. The difference here is that the modulation is generally non-adiabatic -- in fact the modulation frequency $\Omega$ is larger than the starting CCW badwidth $4J$. However, the \textit{time-averaged} group velocity is proportional to $\partial \varepsilon(k) / \partial k$, and so approximately zero for all values of $k$. In Fig. \ref{fig1}(c), we present the Floquet band structure for $J = 0.4\Omega$. This value of the coupling constant $J$ is large enough to break the perturbative analysis, and the bands no longer follow the simple form prescribed by the effective coupling. Furthermore, they are never flat inside the entire $k$-space for any value of $A_0$ when the limit $J \ll \Omega$ is not fulfilled.

\begin{figure}
\centering
\includegraphics[width = 0.45\textwidth, trim = 0in 0in 0in 0in, clip = true]{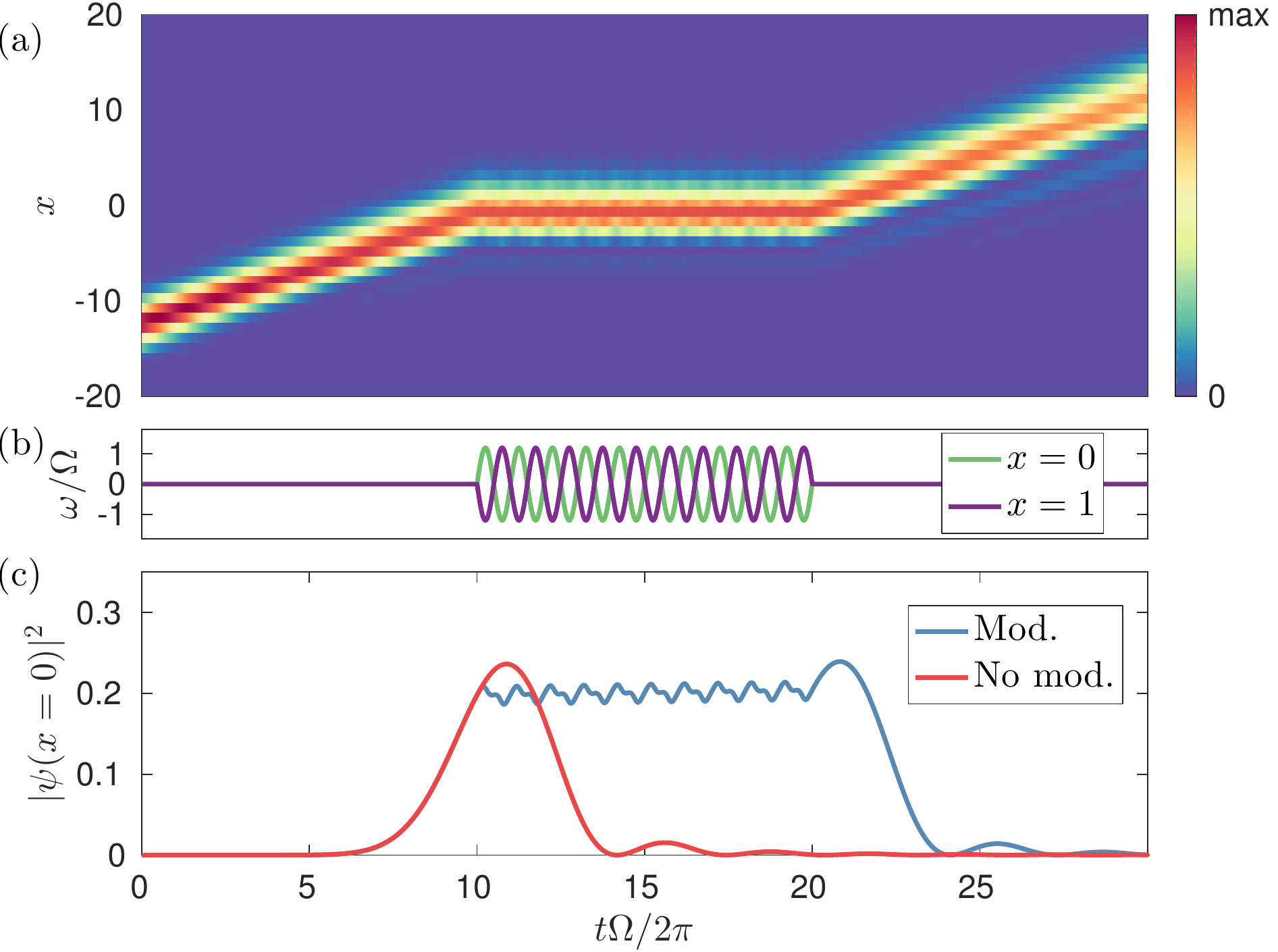}%
 \caption{Wave-packet propagation $|\psi(x, t)|^2$ inside the CCW with $J = 0.1\Omega$, with a modulation $\omega_n = A_0\sin(\Omega t + n\pi)$, $A_0 = 1.2\Omega$, turned on and off at times $10T$ and $20T$, respectively. (a): Real-space propagation. (b): Resonance frequency vs. time for the cavities at positions $x = 0$ and $x = 1$. (c): Field intensity in the cavity at $x = 0$, for a pulse propagating in an unmodulated (red) and a modulated (blue) chain.}
\label{fig:sine}
\end{figure}
In Fig. \ref{fig:sine} we provide an illustration of the localization process by simulating the propagation inside the CCW, starting with a Gaussian wave-packet given by $\psi(x, 0) = \exp(-(x-x_0)^2/(2\sigma_x^2) + ik_0(x-x_0))$. This corresponds, in $k$-space, to a wave-packet centered at $k_0$, with a standard deviation $\sigma_k = 1/\sigma_x$. We set $k_0 = \pi/2$ and $\sigma_x = 2$, so as to have a broad $k$-space distribution. The chain is unmodulated in the beginning and in the end ($A_0 = 0$), while a modulation of the form $A_0\sin(\Omega t + n\pi)$ is applied for $10T < t < 20T$. The evolution of the state, computed by numerically solving eq. (\ref{eqn:cm}) using the Runge-Kutta method, is shown in Fig. \ref{fig:sine}(a). Initially, the wave packet propagates with a group velocity given by $2J\cos(k_0) = 0.2\Omega$. The starting position is $x_0 = -4\pi$, such that at time $t = 10T$ the wave-packet is centered around $x = 0$. Then, during the modulation interval, the state stays perfectly localized around this position. Afterwards, the wave-packet continues its propagation without any distortion caused by the localization. The broadening of the pulse is only due to its broad $k$-space distribution, i.e. the group-velocity dispersion is non-negligible. This is also illustrated in Fig. \ref{fig:sine}(c), where we show the amplitude of the cavity at $x = 0$ vs. time, and compare it versus the case of no modulation (red). As can be seen, the effect of the modulation is only to delay, but not to distort, the evolution of the field amplitude. We stress that Fig. \ref{fig:sine} illustrates the exact time-evolution of the wave-packet, i.e. the approximation of eqs. (2)-(5) was not assumed. 

\begin{figure}
\centering
\includegraphics[width = 0.45\textwidth, trim = 0in 0in 0in 0in, clip = true]{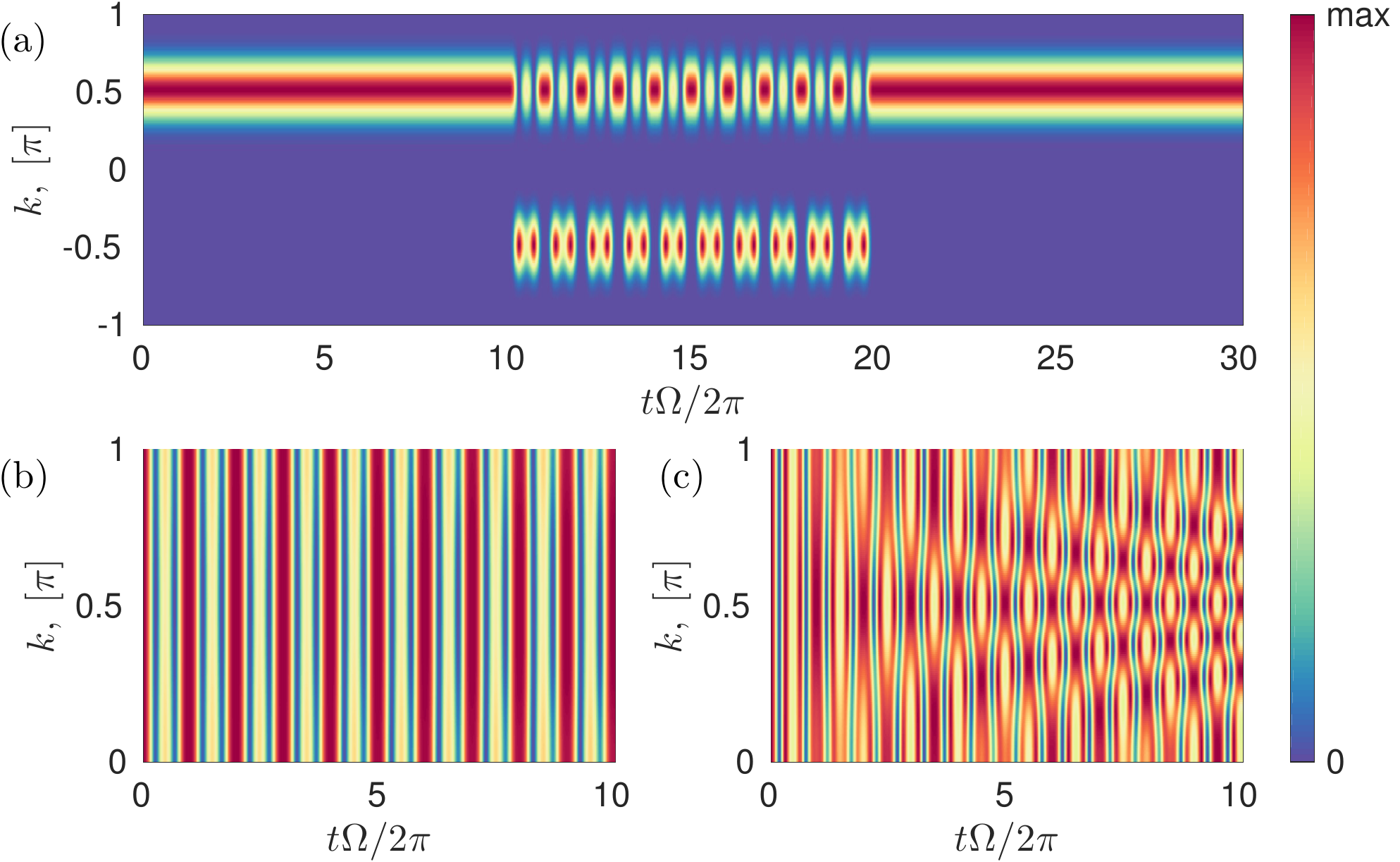}
 \caption{(a): Fourier-space time evolution $|\psi(k, t)|^2$ for the wave-packet and modulation scheme of Fig. \ref{fig:sine}. (b): Time evolution of a starting state $\psi(k, 0) = 1$ for $k = [0, \pi]$, $\psi(k, 0) = 0$ otherwise, under the modulation $\omega_n(t) = A_0 \sin(\Omega t + n\pi)$ for $2A_0/\Omega = 2.4$. (c): Same as (b), for $2A_0/\Omega = 4$.}
\label{fig:sine_k}
\end{figure}

The dynamic simulation shown in Fig. \ref{fig:sine} perfectly confirms the localization expected from the perturbative result above. It is important to note that the latter only holds true for time periods that are an integer multiple of $T$. Specifically, when the effective coupling constant is $J^{(1)} = 0$, the state $\psi(x, t)$ at $t = T$ has to return to its starting value at $t = 0$, but there is no particular constraint on its evolution for intermediate times, which is sometimes referred to as the micromotion of a modulated system \cite{Goldman2014}. The small oscillations of the blue curve of Fig. \ref{fig:sine}(c) are thus expected, and the effect of this micromotion is illustrated further in Fig. \ref{fig:sine_k}(a), where we plot the $k$-space distribution of the state of Fig. \ref{fig:sine} as a function of time. During the time intervals with no modulation, the Bloch momentum is conserved, and $|\psi(k, t)|^2$ stays constant. The modulation, however, has spatial periodicity $a = 2$, and thus conserves $k$ only modulo $\pi$. Consequently, we observe complete, oscillating transfer of energy between the $k$ and $k + \pi$ components of the state during the modulation, with the state returning to its starting value for every $t$ that is integer multiple of $T$. 

Note that in Fig. \ref{fig:sine_k}, we show the full Brillouin zone of width $2\pi$ corresponding to the unmodulated waveguide. In the presence of modulation, the spatial period is $a = 2$, and the Brillouin zone is thus twice smaller. However, plotting with respect to the Brillouin zone of the unmodulated system allows us to clearly exhibit the oscillations between the $k$ and the $k + \pi$ components for the modes in the underlying waveguide. These oscillations can be accounted for analytically. Calling $\alpha_k = \sum_n e^{ikn} \alpha_n$ and $\bm{\alpha}_k = (\alpha_k, \alpha_{k + \pi})$, the equations of motion in reciprocal space are $i \dot{\bm{\alpha}_k} = \mathcal{H}(k)\bm{\alpha}_k$, with
\begin{equation}
\mathcal{H}(k) = \begin{pmatrix}
2J\cos(k) & A_0 \sin(\Omega t) \\
A_0 \sin(\Omega t) & - 2J\cos(k)
\end{pmatrix},
\label{eqn:Hk_sine}
\end{equation}
as we show in the Supplementary Information \cite{SM}. This expression is exact. The Hamiltonian of eq. (\ref{eqn:Hk_sine}) can be identified with that of a two-level system driven by an external sinusoidal electric field of the form $E_0\sin(\Omega t) e^{i\omega_f t}$. The diagonal terms in this analogy define the detuning between $\omega_f$ and the transition frequency $\omega_r$ of the two-level system. On resonance, i.e. for $k = \pi/2$, the dynamics of eq. (\ref{eqn:Hk_sine}) can be solved analytically by defining the \textit{Rabi angle} (see e.g. \cite{Shore2011})
\begin{equation}
\mathcal{A}(t) = \int_{0}^{t} A_0 \sin(\Omega t') \mathrm{d}t' = \frac{2A_0}{\Omega}\sin^2\left(\frac{\Omega t}{2}\right).
\label{eqn:rabi}
\end{equation}
For a starting state $\psi(\pi/2, t = 0) = 1$, $\psi(-\pi/2, t = 0) = 0$, the time evolution is given by $\psi(k = \pi/2, t) = \cos(\mathcal{A}(t))$, $\psi(k = -\pi/2, t) = -i\sin(\mathcal{A}(t))$. At the end of a modulation cycle, we always have $\mathcal{A}(t) = 0$, which means that the $k = \pi/2$ components of the pulse return to their starting amplitude for \textit{every} modulation amplitude $A_0$. The importance of the choice $\mathcal{J}(2A_0/\Omega) = 0$ for localization can however be revealed by studying the remaining $k$-components. There is no analytic solution to the differential equations corresponding to eq. (\ref{eqn:Hk_sine}) for non-zero diagonal elements, but in Fig. \ref{fig:sine_k}(b)-(c) we show the numerical solution for two values of $A_0$, computed again with the Runge-Kutta method. The starting state is $\psi(k, 0) = 1$ for all $k \in [0, 1]$, and $\psi(k, 0) = 0$ otherwise. In panel (b), where $2A_0/\Omega = 2.4$, the oscillations are independent of $k$. Thus, after one cycle, \textit{all} of the Fourier components return to their starting values, as must be the case for dynamic localization. This is not the case for other modulation amplitudes: an example is shown in panel (c), where $2A_0/\Omega  = 4$. Similarly, non-trivial $k$-dependence is observed for all amplitudes for which $\mathcal{J}(2A_0/\Omega) = 0$ does not hold. 

\begin{figure}
\centering
\includegraphics[width = 0.5\textwidth, trim = 0in 0in 0in 0in, clip = true]{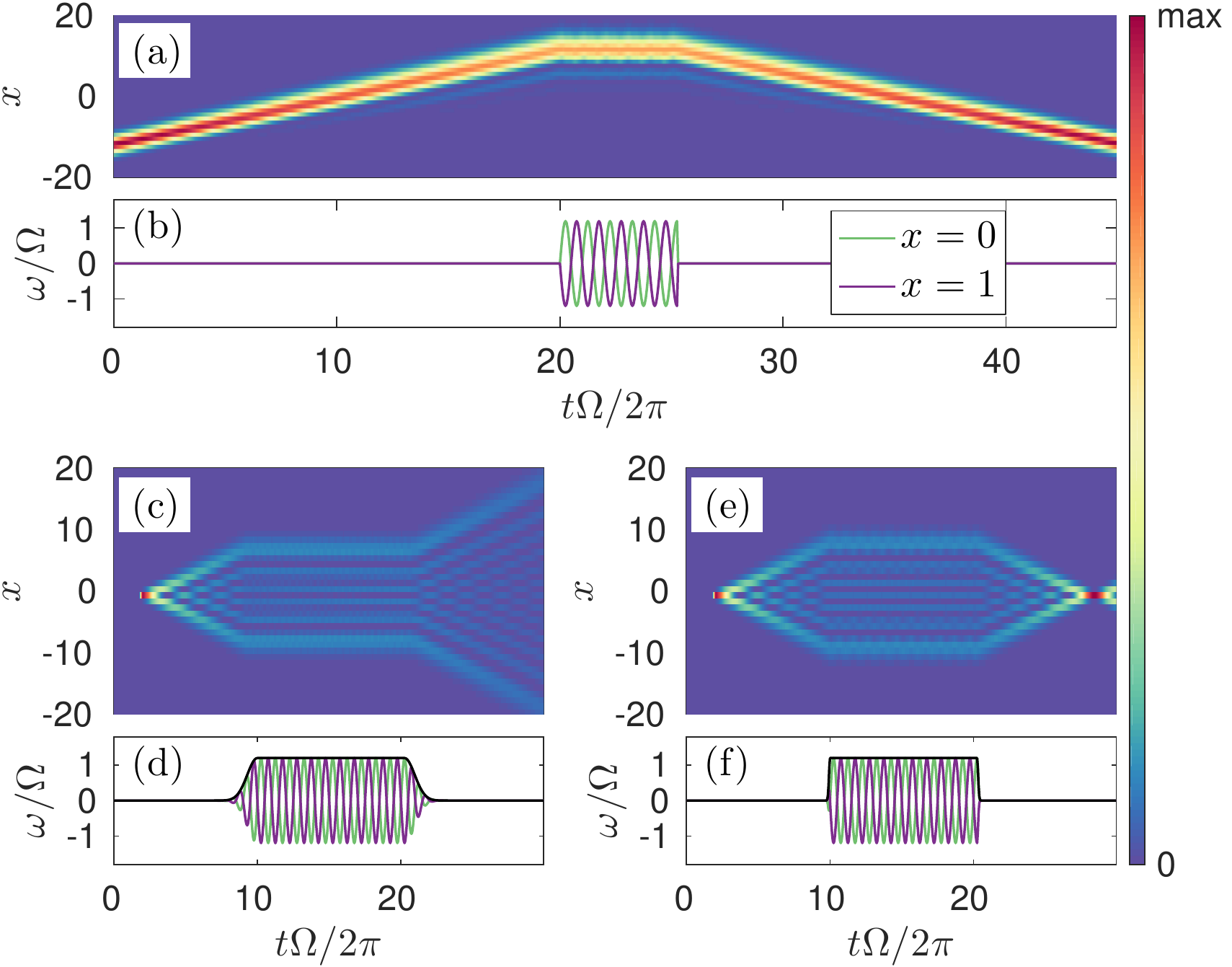}%
 \caption{(a): Propagation of the same starting wave packet as in Fig. \ref{fig:sine}, but with the modulation applied in the interval $20T < t < 25.3T$ as shown in (b). (c): Light from a Gaussian source at $x = 0$ under a localizing modulation that is adiabatically turned on and off, as shown in (d), where the black line is the envelope function $A(t)$. (e)-(f): Same as (c)-(d), but the width $t_w$ of eq. (\ref{eqn:Aenv}) is ten times shorter.}
\label{fig:sine_tr}
\end{figure}

Interestingly, the complete transfer between $k$ and $k + \pi$ components results in complete time-reversal of any arbitrary wave-packet, in the sense that 
\begin{equation}
|\psi(x, -t)|^2  = |\bar{\psi}(x, t)|^2, 
\label{eqn:tr}
\end{equation}
where $\bar{\psi}(k, t) = \psi(k + \pi, t)$, and evolution in an unmodulated CCW is assumed (see the Supplementary Information \cite{SM}). Thus, it is possible to start with a light pulse in a static CCW, apply a specific modulation for a finite amount of time, and return to a static system containing a time-reversed copy of the starting pulse. This is illustrated in Fig. \ref{fig:sine_tr}. In panel (a), we plot the propagation of the same starting pulse as in Fig. \ref{fig:sine}(a), but now we turn on the modulation at $t = 20T$ and then turn it off at $t = 25.3T$. The time-reversal after the modulation is evident: the pulse reverses direction and re-focuses back into its starting form. This time-reversal paradigm is related to the ones discussed in Ref. \cite{Chumak2010, Sivan2011}, and, compared to previous works demonstrating time-reversal in a chain of coupled cavities \cite{Yanik2004, Yuan2016}, has the advantage that the final structure is the same as the starting one -- the static underlying CCW. Beyond revealing the possibility for complete time reversal, another implication of eq. (\ref{eqn:Hk_sine}) is that the requirement $J/\Omega \ll 1$, which we have thus far assumed, can be relaxed for some practical applications. This condition is needed for \textit{complete} localization, in which any arbitrary state returns to its starting amplitude after a full modulation cycle. However, if the starting state is a pulse centered around $k = \pi/2$ of width $\Delta k$, the condition can be modified to $J\cos(\pi/2 + \Delta k/2) \ll \Omega$. This is further illustrated in the Supplementary Information \cite{SM}. 

Finally, we turn our attention to the way the modulation is switched on and off. The modulation schemes of Fig. \ref{fig:sine_tr}(b) and Fig. \ref{fig:sine}(b) have a discontinuity in $\omega_n(t)$ and in $\mathrm{d}\omega_n(t)/\mathrm{d}t$, respectively, which is not realistic. Thus, in the modulation scheme $\omega_n(t) = A(t)\sin(\Omega t + n\pi)$, we study the effect when $A(t)$ is smoothly turned on and off. In this case, the solution above for $k = \pi/2$ is still valid, provided that the Rabi Angle of eq. (\ref{eqn:rabi}) is computed including the envelope function $A(t)$. In the Supplementary Information \cite{SM} we show that, for an adiabatic switching of the modulation with respect to $\Omega$, the Rabi angle $\mathcal{A}(t)$ always goes to zero at the end of the modulation, when the amplitude $A(t)$ goes to zero. This means that the state at $k = \pi/2$ returns to its starting amplitude regardless of the shape of the envelope. Furthermore, as discussed earlier, for a fixed-amplitude modulation with $2A_0/\Omega = 2.4$, the oscillations of the $k$-components do not depend on $k$. This altogether means that complete dynamic localization with a tunable delay time can be achieved by adiabatically turning on the amplitude from zero to $1.2\Omega$ and back. This is illustrated in Fig. \ref{fig:sine_tr}(c)-(f), where we use a modulation envelope of the form
\begin{equation}
A(t) = 
  \begin{cases}
    A_0 \exp\left(-(t-t_1)^2/t_w^2\right) & \quad t \le t_1\\
	A_0 & \quad t_1 < t \le t_2 \\
	A_0 \exp\left(-(t-t_2)^2/t_w^2\right) & \quad t_2 < t\\
  \end{cases}
  \label{eqn:Aenv}
\end{equation}
This time, the starting state $\psi(x, 0)$ is zero everywhere, and we have placed a source with frequency $\omega_s = \omega_0$ and a Gaussian pulse-shape $S(t) = \exp(-(t - 2T)^2/T^2)$ in the cavity at $x = 0$. In Fig. \ref{fig:sine_tr}(c)-(d), we show the propagation and localization of the emitted light, with $t_1 = 10T$, $t_2 = 20.25T$, $t_w = 2\pi/\Omega$. Because the envelope is adiabatic for this value of $t_w$, the state returns to its starting form for any arbitrary interval determined by $t_1$ and $t_2$. This shows that precise timing control of the modulation is not needed for an optical delay line. On the other hand, this also means that, in order to achieve the time-reversal scheme of Fig. \ref{fig:sine_tr}(a), the envelope must be non-adiabatic, and so precise control is needed. The general condition, as per eq. (\ref{eqn:tr}), is that after the modulation, all the $k$-components of the state are transferred to $k + \pi$. Thus, the precise parameters for a particular system have to be tuned by solving the time evolution as prescribed by eq. (\ref{eqn:Hk_sine}). An example is illustrated in Fig. \ref{fig:sine_tr}(e)-(f), where $t_1$ and $t_2$ are the same as in panels (a)-(b), but the temporal on/off width is ten times smaller, $t_w = 0.2\pi/\Omega$. Complete time reversal is achieved for this particular set of $t_1$ and $t_2$, and the light is re-focused at the original source position. 

In conclusion, we have shown a scheme for dynamic localization that is particularly relevant to optical systems, and we have revealed its connection to a time-reversal operation, achieved when the modulation is switched off in a well-controlled way. We focused on a coupled-cavity waveguide for simplicity, which can be implemented experimentally for example through fibre-based or integrated cavities and electro-optic modulators \cite{SM, Reed2010, Xu2005, Okamura1991, Asano2017}. Furthermore, we expect that an expression similar to eq. (\ref{eqn:Hk_sine}) -- and all the subsequent results -- can also be derived for a standard, continuous waveguide. We discuss this further in the Supplementary Information \cite{SM}, and suggest an experimental realization in a waveguide modulated using the schemes of e.g. \cite{Leonard2002, Beggs2012} or \cite{Lira2012, Tzuang2014}. In short, our proposal falls within the state-of-the-art feasibility of several experimental paradigms, and has an array of possible applications in photonic technologies.

\begin{acknowledgments}
This work was supported by the Swiss National Science Foundation through Project N\textsuperscript{\underline{o}} P2ELP2\_165174, and the US Air Force Office of Scientific Research FA9550-17-1-0002. 
\end{acknowledgments}


%

\end{document}